\documentclass[12pt]{article}
\usepackage{amsmath,amssymb,amsthm,amsxtra,overpic,bbm,bm,epsfig,ulem}
\usepackage{color}
\usepackage{cite}

\usepackage{hyperref}
\usepackage{url}

\def\thefootnote{\fnsymbol{footnote}}

\begin{document}

\vspace{0.2cm}

\begin{center}
{\large\bf Rephasing invariants of CP violation for heavy and light Majorana neutrinos}
\end{center}

\vspace{0.2cm}

\begin{center}
{\bf Zhi-zhong Xing$^{1,2,3}$}
\footnote{E-mail: xingzz@ihep.ac.cn}
\\
{\small $^{1}$Institute of High Energy Physics, Chinese Academy of Sciences,
Beijing 100049, China} \\
{\small $^{2}$School of Physical Sciences,
University of Chinese Academy of Sciences, Beijing 100049, China} \\
{$^{3}$Center of High Energy Physics, Peking University, Beijing 100871, China}
\end{center}

\vspace{1cm}

\begin{abstract}
In the canonical seesaw mechanism, the strengths of charged-current interactions
for light and heavy Majorana neutrinos are described respectively by the
$3\times 3$ matrices $U$ and $R$ that are correlated with each other via
the exact seesaw relation and the unitarity condition. We write out the
Majorana-type invariants of CP violation of $R$ and $U$, which are
insensitive to redefining the phases of three charged-lepton fields; and
the Dirac-type invariants of CP violation of $R$ and $U$ that are insensitive
to the rephasing of both the charged-lepton fields and the neutrino fields.
Such invariants are explicitly calculated with the help of a full
Euler-like block parametrization of the seesaw flavor structure containing
nine active-sterile flavor mixing angles and six independent CP-violating phases,
and their corresponding roles in the CP-violating asymmetries of three heavy
Majorana neutrino decays and in the flavor oscillations of three light Majorana
neutrinos are briefly discussed. We point out that similar rephasing invariants
arising from the interplay between $R$ and $U$ may also manifest themselves in
a variety of lepton-flavor-violating and lepton-number-violating processes.
\end{abstract}

\newpage

\def\thefootnote{\arabic{footnote}}
\setcounter{footnote}{0}

\section{Introduction}

Given the canonical seesaw mechanism~\cite{Minkowski:1977sc,Yanagida:1979as,
GellMann:1980vs,Glashow:1979nm,Mohapatra:1979ia} and the associated
leptogenesis mechanism~\cite{Fukugita:1986hr} as the most natural
and economical ``stone" beyond the standard model (SM) to kill ``two birds"
--- the origin of tiny masses for three active neutrinos and the dynamics
of cosmic baryogenesis, we are certainly interested in studying various
aspects of their theoretical properties and phenomenological consequences.
CP violation is exactly one of the most important issues in this connection.

The nontrivial phases responsible for CP violation in the light and heavy
Majorana neutrino sectors of the seesaw mechanism are hidden respectively in
the $3 \times 3$ Pontecorvo-Maki-Nakagawa-Sakata (PMNS) flavor mixing
matrix $U$~\cite{Pontecorvo:1957cp,Maki:1962mu,Pontecorvo:1967fh} and
its counterpart $R$ which characterize the strengths of leptonic weak
charged-current interactions of the form~\cite{Xing:2007zj,Xing:2011ur}
\begin{eqnarray}
-{\cal L}^{}_{\rm cc} = \frac{g}{\sqrt{2}} \hspace{0.1cm}
\overline{\big(\begin{matrix} e & \mu & \tau\end{matrix}\big)^{}_{\rm L}}
\hspace{0.1cm} \gamma^\mu \left[ U \left( \begin{matrix} \nu^{}_{1}
\cr \nu^{}_{2} \cr \nu^{}_{3} \cr\end{matrix}
\right)^{}_{\hspace{-0.08cm} \rm L}
+ R \left(\begin{matrix} N^{}_4 \cr N^{}_5 \cr N^{}_6
\cr\end{matrix}\right)^{}_{\hspace{-0.08cm} \rm L} \hspace{0.05cm} \right]
W^-_\mu + {\rm h.c.} \; ,
\label{1}
\end{eqnarray}
where the fields of charged leptons ($e, \mu, \tau$), light Majorana neutrinos
($\nu^{}_i$ for $i = 1, 2, 3$) and heavy Majorana neutrinos ($N^{}_j$
for $j = 4, 5, 6$) are all in the mass basis. The flavor parameters of
$U$ and $R$ are correlated with each other via both the unitarity condition
$U U^\dagger + R R^\dagger = I$ and the exact seesaw relation
\begin{eqnarray}
U \left(\begin{matrix}
m^{}_1 & 0 & 0 \cr
0 & m^{}_2 & 0 \cr
0 & 0 & m^{}_3 \cr \end{matrix} \right) U^T +
R \left(\begin{matrix}
M^{}_4 & 0 & 0 \cr
0 & M^{}_5 & 0 \cr
0 & 0 & M^{}_6 \cr \end{matrix} \right) R^T =
\left(\begin{matrix}
0 & 0 & 0 \cr
0 & 0 & 0 \cr
0 & 0 & 0 \cr \end{matrix} \right)
\label{2}
\end{eqnarray}
with $m^{}_i$ and $M^{}_j$ being the respective masses of light and heavy
neutrinos (for $i = 1, 2, 3$ and $j = 4, 5, 6$). A free redefinition of
the phases of three charged-lepton fields is allowed, but this will affect
the phase structures of $U$ and $R$. In comparison, a common phase of
the three light Majorana neutrino fields $\nu^{}_i$ (or the three heavy
Majorana neutrino fields $N^{}_j$) can also be arbitrarily chosen. Such
degrees of arbitrariness imply that the matrix elements of $U$ and $R$
are not rephasing invariant; namely, some phase parameters associated
with $U^{}_{\alpha i}$ or $R^{}_{\alpha j}$ (for $\alpha = e, \mu, \tau$;
$i = 1, 2, 3$ and $j = 4, 5, 6$) are convention-dependent but irrelevant
to any physical effects of leptonic CP violation.

Now that the observable quantities of CP violation in all the weak-interaction
processes must be rephasing invariant, it makes a lot of sense to construct
the rephasing invariants of CP violation in a given theory and then link
them to the explicit observable effects. In this regard the univeral
Jarlskog invariant proves to be very useful for describing the strength
of CP violation in the quark sector of the SM in a basis-independent
way~\cite{Jarlskog:1985ht,Wu:1985ea}. A similar rephasing invariant of this
kind, denoted as ${\cal J}^{}_\nu$, is even more useful in the lepton sector
as it will uniquely measure the strength of CP violation in neutrino
oscillations if the $3 \times 3$ PMNS matrix $U$ is assumed to be
unitary~\cite{Barger:1980jm,Cheng:1986in}.

But $U$ has to deviate slightly from its unitarity limit $U^{}_0$ in the
canonical seesaw mechanism, as a straightforward consequence of small
active-sterile flavor mixing effects characterized by nonzero $R$. In
other words, $R = {\bf 0}$ would lead to $m^{}_i = 0$ as can be easily
seen from Eq.~(\ref{2}), and hence $U$ would lose its physical meaning
in this extreme case. Taking account of the non-unitarity of $U$, one may
define nine distinctive Jarlskog-like invariants of CP violation as
\begin{eqnarray}
{\cal J}^{i i^\prime}_{\alpha \beta} \equiv {\rm Im}
\left(U^{}_{\alpha i} U^{}_{\beta i^\prime} U^*_{\alpha i^\prime}
U^*_{\beta i}\right) \; ,
\label{3}
\end{eqnarray}
where the Greek subscripts $(\alpha, \beta)$ run over
$(e, \mu, \tau)$, and the Latin subscripts $(i, i^\prime)$ run
over $(1, 2, 3)$. The leading terms of ${\cal J}^{i i^\prime}_{\alpha \beta}$
are found to be all equal to $\pm {\cal J}^{}_\nu$
in the $U \to U^{}_0$ approximation~\cite{Xing:2007zj,Xing:2011ur}, thanks to
the smallness of active-sterile flavor mixing as constrained by a variety of
precision electroweak and flavor measurements~\cite{Antusch:2006vwa,
Blennow:2016jkn,Blennow:2023mqx,Xing:2024gmy}. As for CP violation in
neutrino-antineutrino oscillations, the corresponding Jarlskog-like invariants
can be defined as~\cite{Luo:2011mm,Xing:2013ty,Xing:2013woa,Wang:2021rsi}
\begin{eqnarray}
{\cal V}^{i i^\prime}_{\alpha \beta} \equiv {\rm Im}
\left(U^{}_{\alpha i} U^{}_{\beta i} U^*_{\alpha i^\prime}
U^*_{\beta i^\prime}\right) \; ,
\label{4}
\end{eqnarray}
which are insensitive to a common phase of the three light Majorana neutrino
fields but sensitive to their relative phase differences. That is to say,
${\cal J}^{i i^\prime}_{\alpha \beta}$ and ${\cal V}^{i i^\prime}_{\alpha \beta}$
measure the rephasing-invariant effects of leptonic CP violation with
lepton number conservation and lepton number nonconservation, respectively.

Along this line of thought, let us define the analogous rephasing invariants
of CP violation for heavy Majorana neutrinos in terms of the matrix elements
of $R$ in the canonical seesaw framework:
\begin{eqnarray}
&& {\cal X}^{j j^\prime}_{\alpha \beta} \equiv
{\rm Im} \left(R^{}_{\alpha j} R^{}_{\beta j^\prime} R^*_{\alpha j^\prime}
R^*_{\beta j}\right) \; ,
\nonumber \\
&& {\cal Z}^{j j^\prime}_{\alpha \beta} \equiv
{\rm Im} \left(R^{}_{\alpha j} R^{}_{\beta j} R^*_{\alpha j^\prime}
R^*_{\beta j^\prime}\right) \; , \hspace{1.5cm}
\label{5}
\end{eqnarray}
where the Greek subscripts $(\alpha, \beta)$ run over $(e, \mu, \tau)$,
and the Latin subscripts $(j, j^\prime)$ run over $(4, 5, 6)$.
These rephasing-invariant quantities, which have not been considered
before, measure the sizes of CP-violating asymmetries between the heavy
Majorana neutrino decays and their CP-conjugated processes. The explicit
analytical expressions of ${\cal X}^{j j^\prime}_{\alpha \beta}$ and
${\cal Z}^{j j^\prime}_{\alpha \beta}$ will be presented with the help of
a full Euler-like block parametrization of the seesaw flavor structure
that has been proposed in Refs.~\cite{Xing:2007zj,Xing:2011ur}.

It is worth pointing out that the interplay between $U$ and $R$ in some
lepton-flavor-violating (LFV) and lepton-number-violating (LNV) processes
mediated by both $\nu^{}_i$ and $N^{}_j$ allows us to define two new types
of rephasing invariants of the forms
\begin{eqnarray}
&& \mathbb{X}^{i j}_{\alpha \beta} \equiv
U^{}_{\alpha i} R^{}_{\beta j} R^*_{\alpha j} U^*_{\beta i} \; ,
\nonumber \\
&& \mathbb{Z}^{i j}_{\alpha \beta} \equiv
U^{}_{\alpha i} R^{*}_{\beta j} R^*_{\alpha j} U^{}_{\beta i} \; ,
\hspace{1.5cm}
\label{6}
\end{eqnarray}
where $i$ and $j$ run respectively over $(1, 2, 3)$ and $(4, 5, 6)$ for
$\alpha, \beta = e, \mu, \tau$. We are going to illustrate the interesting
roles of $\mathbb{X}^{i j}_{\alpha \beta}$ and $\mathbb{Z}^{i j}_{\alpha \beta}$
in the LFV radiative decays of charged leptons and the LNV neutrinoless
double-beta ($0\nu 2\beta$) decays of some nuclei.

The remaining parts of this paper are organized as follows. In section 2, we
calculate the rephasing invariants ${\cal X}^{j j^\prime}_{\alpha \beta}$
and ${\cal Z}^{j j^\prime}_{\alpha \beta}$ in terms of the nine active-sterile
flavor mixing angles and six independent CP-violating phases of $R$, and discuss
their roles in the CP-violating asymmetries of three heavy Majorana
neutrino decays. Section 3 is devoted to examining to what extent the rephasing
invariants ${\cal J}^{i i^\prime}_{\alpha \beta}$ deviate from $\pm {\cal J}^{}_\nu$
in neutrino oscillations. Finally, we make some brief comments on the invariants
$\mathbb{X}^{i j}_{\alpha \beta}$ and $\mathbb{Z}^{i j}_{\alpha \beta}$ associated
with two typical LFV and LNV processes in section 4.

\section{The expressions of ${\cal X}^{j j^\prime}_{\alpha \beta}$
and ${\cal Z}^{j j^\prime}_{\alpha \beta}$}

As the $3 \times 3$ flavor mixing matrices $U$ and $R$ in Eq.~(\ref{1}) are actually
two sub-matrices of a $6 \times 6$
unitary matrix $\mathbb{U}$ used to diagonalize the flavor basis of all the
six (three active and three sterile) neutrino fields in the canonical seesaw
framework, they are transparently correlated with each other when $\mathbb{U}$
is fully parameterized in terms of the Euler-like rotation angles and
phase angles~\cite{Xing:2007zj}. Denoting $U = A U^{}_0$ with
$A$ being a $3 \times 3$ lower triangular matrix and $U^{}_0$ being a
$3 \times 3$ unitary matrix, we obtain the explicit expressions of $A$, $R$
and $U^{}_0$ as follows~\cite{Xing:2011ur}:
\begin{eqnarray}
A \hspace{-0.2cm} & = & \hspace{-0.2cm}
\left( \begin{matrix} c^{}_{14} c^{}_{15} c^{}_{16} & 0 & 0
\cr \vspace{-0.45cm} \cr
\begin{array}{l} -c^{}_{14} c^{}_{15} \hat{s}^{}_{16} \hat{s}^*_{26} -
c^{}_{14} \hat{s}^{}_{15} \hat{s}^*_{25} c^{}_{26} \\
-\hat{s}^{}_{14} \hat{s}^*_{24} c^{}_{25} c^{}_{26} \end{array} &
c^{}_{24} c^{}_{25} c^{}_{26} & 0 \cr \vspace{-0.45cm} \cr
\begin{array}{l} -c^{}_{14} c^{}_{15} \hat{s}^{}_{16} c^{}_{26} \hat{s}^*_{36}
+ c^{}_{14} \hat{s}^{}_{15} \hat{s}^*_{25} \hat{s}^{}_{26} \hat{s}^*_{36} \\
- c^{}_{14} \hat{s}^{}_{15} c^{}_{25} \hat{s}^*_{35} c^{}_{36} +
\hat{s}^{}_{14} \hat{s}^*_{24} c^{}_{25} \hat{s}^{}_{26}
\hat{s}^*_{36} \\
+ \hat{s}^{}_{14} \hat{s}^*_{24} \hat{s}^{}_{25} \hat{s}^*_{35}
c^{}_{36} - \hat{s}^{}_{14} c^{}_{24} \hat{s}^*_{34} c^{}_{35}
c^{}_{36} \end{array} &
\begin{array}{l} -c^{}_{24} c^{}_{25} \hat{s}^{}_{26} \hat{s}^*_{36} -
c^{}_{24} \hat{s}^{}_{25} \hat{s}^*_{35} c^{}_{36} \\
-\hat{s}^{}_{24} \hat{s}^*_{34} c^{}_{35} c^{}_{36} \end{array} &
c^{}_{34} c^{}_{35} c^{}_{36} \cr \end{matrix} \right) \; ,
\nonumber \\
R \hspace{-0.2cm} & = & \hspace{-0.2cm}
\left( \begin{matrix} \hat{s}^*_{14} c^{}_{15} c^{}_{16} &
\hat{s}^*_{15} c^{}_{16} & \hat{s}^*_{16} \cr \vspace{-0.45cm} \cr
\begin{array}{l} -\hat{s}^*_{14} c^{}_{15} \hat{s}^{}_{16} \hat{s}^*_{26} -
\hat{s}^*_{14} \hat{s}^{}_{15} \hat{s}^*_{25} c^{}_{26} \\
+ c^{}_{14} \hat{s}^*_{24} c^{}_{25} c^{}_{26} \end{array} & -
\hat{s}^*_{15} \hat{s}^{}_{16} \hat{s}^*_{26} + c^{}_{15}
\hat{s}^*_{25} c^{}_{26} & c^{}_{16} \hat{s}^*_{26} \cr \vspace{-0.45cm} \cr
\begin{array}{l} -\hat{s}^*_{14} c^{}_{15} \hat{s}^{}_{16} c^{}_{26}
\hat{s}^*_{36} + \hat{s}^*_{14} \hat{s}^{}_{15} \hat{s}^*_{25}
\hat{s}^{}_{26} \hat{s}^*_{36} \\ - \hat{s}^*_{14} \hat{s}^{}_{15}
c^{}_{25} \hat{s}^*_{35} c^{}_{36} - c^{}_{14} \hat{s}^*_{24}
c^{}_{25} \hat{s}^{}_{26}
\hat{s}^*_{36} \\
- c^{}_{14} \hat{s}^*_{24} \hat{s}^{}_{25} \hat{s}^*_{35}
c^{}_{36} + c^{}_{14} c^{}_{24} \hat{s}^*_{34} c^{}_{35} c^{}_{36}
\end{array} &
\begin{array}{l} -\hat{s}^*_{15} \hat{s}^{}_{16} c^{}_{26} \hat{s}^*_{36}
- c^{}_{15} \hat{s}^*_{25} \hat{s}^{}_{26} \hat{s}^*_{36} \\
+c^{}_{15} c^{}_{25} \hat{s}^*_{35} c^{}_{36} \end{array} &
c^{}_{16} c^{}_{26} \hat{s}^*_{36} \cr \end{matrix} \right) \; ,
\hspace{0.5cm}
\label{7}
\end{eqnarray}
where $c^{}_{ij} \equiv \cos\theta^{}_{ij}$, $s^{}_{ij} \equiv \sin\theta^{}_{ij}$
and $\hat{s}^{}_{ij} \equiv s^{}_{ij} e^{{\rm i}\delta^{}_{ij}}$ with
$\theta^{}_{ij}$ and $\delta^{}_{ij}$ being the active-sterile flavor mixing
angles and the corresponding CP-violating phases (for $i = 1, 2, 3$ and
$j = 4, 5, 6$); and
\begin{eqnarray}
U^{}_0 = \left( \begin{matrix} c^{}_{12} c^{}_{13} & \hat{s}^*_{12}
c^{}_{13} & \hat{s}^*_{13} \cr
-\hat{s}^{}_{12} c^{}_{23} -
c^{}_{12} \hat{s}^{}_{13} \hat{s}^*_{23} & c^{}_{12} c^{}_{23} -
\hat{s}^*_{12} \hat{s}^{}_{13} \hat{s}^*_{23} & c^{}_{13}
\hat{s}^*_{23} \cr
\hat{s}^{}_{12} \hat{s}^{}_{23} - c^{}_{12}
\hat{s}^{}_{13} c^{}_{23} & -c^{}_{12} \hat{s}^{}_{23} -
\hat{s}^*_{12} \hat{s}^{}_{13} c^{}_{23} & c^{}_{13} c^{}_{23}
\cr \end{matrix} \right) \; , \hspace{0.4cm}
\label{8}
\end{eqnarray}
in which $c^{}_{i i^\prime} \equiv \cos\theta^{}_{i i^\prime}$,
$s^{}_{i i^\prime} \equiv \sin\theta^{}_{i i^\prime}$
and $\hat{s}^{}_{i i^\prime} \equiv s^{}_{i i^\prime}
e^{{\rm i}\delta^{}_{i i^\prime}}$ with $\theta^{}_{i i^\prime}$ and
$\delta^{}_{i i^\prime}$ being the active flavor mixing angles and
the corresponding CP-violating phases (for $i i^\prime = 12, 13, 23$).
Note, however, that the six flavor parameters of $U^{}_0$ and the
three light Majorana neutrino masses are {\it derivational} in the sense
that they can all be derived from the {\it original} seesaw flavor
parameters emerging from the seesaw Lagrangian
(i.e., the three heavy Majorana neutrino masses,
nine active-sterile flavor mixing angles and six independent CP-violating
phases of $A$ and $R$), as can be seen from the exact seesaw realtion
in Eq.~(\ref{2})
\footnote{The explicit analytical expressions of $m^{}_i$,
$\theta^{}_{i i^\prime}$ and $\delta^{}_{i i^\prime}$ (for $i = 1, 2, 3$
and $i i^\prime = 12, 13, 23$) in terms of $M^{}_j$, $\theta^{}_{ij}$
and $\delta^{}_{ij}$ (for $i = 1, 2, 3$ and $j = 4, 5, 6$) are very
tedious and can be found in Refs.~\cite{Xing:2024xwb,Xing:2024gmy}.}.

A recent careful global analysis of the currently available electroweak
precision measurements and neutrino oscillation data has provided us with
rather stringent numerical bounds on the possible deviation of
$U U^\dagger = A A^\dagger = I - R R^\dagger$
from $I$~\cite{Blennow:2023mqx,Xing:2024gmy}. Such constraints can
easily be translated into the upper limits on the nine active-sterile
flavor mixing angles:
\begin{eqnarray}
s^{}_{1j} < 0.051 \; , \quad
s^{}_{2j} < 0.0047 \; , \quad
s^{}_{3j} < 0.045 \;
\label{9}
\end{eqnarray}
in the normal neutrino mass ordering ($m^{}_1 < m^{}_2 < m^{}_3$) case;
or
\begin{eqnarray}
s^{}_{1j} < 0.053 \; , \quad
s^{}_{2j} < 0.0045 \; , \quad
s^{}_{3j} < 0.040 \;
\label{10}
\end{eqnarray}
in the inverted neutrino mass ordering ($m^{}_3 < m^{}_1 < m^{}_2$) case,
where $j = 4, 5, 6$. One is therefore allowed to safely expand the elements of $A$
and $R$ as Maclaurin's series of the form
\begin{eqnarray}
A =
I - \left( \begin{matrix} a^{}_{11} & 0 & 0 \cr
a^{}_{21} & a^{}_{22} & 0 \cr
a^{}_{31} & a^{}_{32} & a^{}_{33} \cr \end{matrix}
\right) + {\cal O}\left(s^4_{ij}\right) \; ,
\quad
R =
\left( \begin{matrix} \hat{s}^*_{14} &
\hat{s}^*_{15} & \hat{s}^*_{16} \cr
\hat{s}^*_{24} & \hat{s}^*_{25} &
\hat{s}^*_{26} \cr
\hat{s}^*_{34} & \hat{s}^*_{35} &
\hat{s}^*_{36} \cr \end{matrix} \right) + {\cal O}\left(s^3_{ij}\right) \; ,
\label{11}
\end{eqnarray}
where
\begin{eqnarray}
a^{}_{ii} \equiv \frac{1}{2} \left(s^2_{i4} + s^2_{i 5} + s^2_{i 6}\right) \; ,
\quad
a^{}_{i i^\prime} \equiv \hat{s}^*_{i 4} \hat{s}^{}_{i^\prime 4} +
\hat{s}^*_{i 5} \hat{s}^{}_{i^\prime 5} + \hat{s}^*_{i 6} \hat{s}^{}_{i^\prime 6} \;
\label{12}
\end{eqnarray}
for $ii = 11, 22, 33$ and $i i^\prime = 21, 31, 32$. We find that the
leading-order approximation of $R$ represents a good degree of accuracy
for us to calculate the rephasing invariants defined in Eqs.~(\ref{3})---(\ref{6}).

\subsection{The results of ${\cal X}^{j j^\prime}_{\alpha \beta}$}

Given the definition of the rephasing invariants
${\cal X}^{j j^\prime}_{\alpha \beta}$ in Eq.~(\ref{5}) for three heavy
Majorana neutrinos, we find that ${\cal X}^{j j^\prime}_{\alpha \alpha} =
{\cal X}^{j j}_{\alpha \beta} = 0$ and
${\cal X}^{j j^\prime}_{\beta \alpha} = {\cal X}^{j^\prime j}_{\alpha \beta}
= -{\cal X}^{j j^\prime}_{\alpha \beta}$ hold. So it is only needed to calculate
those of ${\cal X}^{j j^\prime}_{\alpha \beta}$ with $\alpha < \beta$ and
$j < j^\prime$. Taking account of the leading-order term of $R$ in
Eq.~(\ref{11}), we immediately arrive at the analytical expressions of nine
distinctive ${\cal X}^{j j^\prime}_{\alpha \beta}$ as follows:
\begin{align}
& {\cal X}^{4 5}_{e \mu} = s^{}_{14} s^{}_{15} s^{}_{24} s^{}_{25}
\sin\left(\alpha^{}_2 - \alpha^{}_1\right) \; ,
\nonumber \\
& {\cal X}^{4 5}_{e \tau} = s^{}_{14} s^{}_{15} s^{}_{34} s^{}_{35}
\sin\left(\alpha^{}_3 - \alpha^{}_1\right) \; ,
\nonumber \\
& {\cal X}^{4 5}_{\mu \tau} = s^{}_{24} s^{}_{25} s^{}_{34} s^{}_{35}
\sin\left(\alpha^{}_3 - \alpha^{}_2\right) \; ; \hspace{1.5cm}
\nonumber \\
& {\cal X}^{4 6}_{e \mu} = s^{}_{14} s^{}_{16} s^{}_{24} s^{}_{26}
\sin\left(\gamma^{}_1 - \gamma^{}_2\right) \; ,
\nonumber \\
& {\cal X}^{4 6}_{e \tau} = s^{}_{14} s^{}_{16} s^{}_{34} s^{}_{36}
\sin\left(\gamma^{}_1 - \gamma^{}_3\right) \; ,
\nonumber \\
& {\cal X}^{4 6}_{\mu \tau} = s^{}_{24} s^{}_{26} s^{}_{34} s^{}_{36}
\sin\left(\gamma^{}_2 - \gamma^{}_3\right) \; ; \hspace{1.5cm}
\nonumber \\
& {\cal X}^{5 6}_{e \mu} = s^{}_{15} s^{}_{16} s^{}_{25} s^{}_{26}
\sin\left(\beta^{}_2 - \beta^{}_1\right) \; ,
\nonumber \\
& {\cal X}^{5 6}_{e \tau} = s^{}_{15} s^{}_{16} s^{}_{35} s^{}_{36}
\sin\left(\beta^{}_3 - \beta^{}_1\right) \; ,
\nonumber \\
& {\cal X}^{5 6}_{\mu \tau} = s^{}_{25} s^{}_{26} s^{}_{35} s^{}_{36}
\sin\left(\beta^{}_3 - \beta^{}_2\right) \; , \hspace{1.5cm}
\label{13}
\end{align}
where the nine CP-violating phases are defined as
\begin{eqnarray}
\alpha^{}_i \equiv \delta^{}_{i 4} - \delta^{}_{i 5} \; , \quad
\beta^{}_i \equiv \delta^{}_{i 5} - \delta^{}_{i 6} \; , \quad
\gamma^{}_i \equiv \delta^{}_{i 6} - \delta^{}_{i 4} \;
\label{14}
\end{eqnarray}
with $\alpha^{}_i + \beta^{}_i + \gamma^{}_i = 0$ for $i = 1, 2, 3$.
That is why only six of the above nine phase parameters are independent
and serve as the original CP-violating phases in the canonical seesaw
mechanism. It should be noted that the invariants
${\cal X}^{j j^\prime}_{\alpha \beta}$ are insensitive to the rephasing
of both the charged-lepton fields and the heavy Majorana neutrino fields.
Therefore, the phase combinations appearing in Eq.~(\ref{13})
cannot reflect all the LNV properties of CP violation in those heavy
Majorana neutrino decays. Let us proceed to consider the invariants
${\cal Z}^{j j^\prime}_{\alpha \beta}$, which contain some other phase
combinations.

\subsection{The results of ${\cal Z}^{j j^\prime}_{\alpha \beta}$}

The definition of ${\cal Z}^{j j^\prime}_{\alpha \beta}$ in Eq.~(\ref{5})
implies that ${\cal Z}^{j j}_{\alpha \beta} = 0$ and
${\cal Z}^{j j^\prime}_{\beta \alpha} = -{\cal Z}^{j^\prime j}_{\alpha \beta}
= {\cal Z}^{j j^\prime}_{\alpha \beta}$ hold. So it is only necessary to
calculate those of ${\cal Z}^{j j^\prime}_{\alpha \beta}$ with $\alpha \leq \beta$
and $j < j^\prime$. Taking account of the leading-order term of $R$ in
Eq.~(\ref{11}), we obtain the analytical expressions of eighteen
distinctive ${\cal Z}^{j j^\prime}_{\alpha \beta}$ as follows:
\begin{align}
& {\cal Z}^{4 5}_{e \mu} = -s^{}_{14} s^{}_{15} s^{}_{24} s^{}_{25}
\sin\left(\alpha^{}_1 + \alpha^{}_2\right) \; ,
\nonumber \\
& {\cal Z}^{4 5}_{e \tau} = -s^{}_{14} s^{}_{15} s^{}_{34} s^{}_{35}
\sin\left(\alpha^{}_1 + \alpha^{}_3\right) \; ,
\nonumber \\
& {\cal Z}^{4 5}_{\mu \tau} = -s^{}_{24} s^{}_{25} s^{}_{34} s^{}_{35}
\sin\left(\alpha^{}_2 + \alpha^{}_3\right) \; ; \hspace{1.5cm}
\nonumber \\
& {\cal Z}^{4 6}_{e \mu} = -s^{}_{14} s^{}_{16} s^{}_{24} s^{}_{26}
\sin\left(\gamma^{}_1 + \gamma^{}_2\right) \; ,
\nonumber \\
& {\cal Z}^{4 6}_{e \tau} = -s^{}_{14} s^{}_{16} s^{}_{34} s^{}_{36}
\sin\left(\gamma^{}_1 + \gamma^{}_3\right) \; ,
\nonumber \\
& {\cal Z}^{4 6}_{\mu \tau} = -s^{}_{24} s^{}_{26} s^{}_{34} s^{}_{36}
\sin\left(\gamma^{}_2 + \gamma^{}_3\right) \; ; \hspace{1.5cm}
\nonumber \\
& {\cal Z}^{5 6}_{e \mu} = -s^{}_{15} s^{}_{16} s^{}_{25} s^{}_{26}
\sin\left(\beta^{}_1 + \beta^{}_2\right) \; ,
\nonumber \\
& {\cal Z}^{5 6}_{e \tau} = -s^{}_{15} s^{}_{16} s^{}_{35} s^{}_{36}
\sin\left(\beta^{}_1 + \beta^{}_3\right) \; ,
\nonumber \\
& {\cal Z}^{5 6}_{\mu \tau} = -s^{}_{25} s^{}_{26} s^{}_{35} s^{}_{36}
\sin\left(\beta^{}_2 + \beta^{}_3\right) \; ; \hspace{1.5cm}
\label{15}
\end{align}
together with
\begin{align}
& {\cal Z}^{4 5}_{e e} = -s^{2}_{14} s^{2}_{15} \sin 2\alpha^{}_1 \; ,
\nonumber \\
& {\cal Z}^{4 5}_{\mu \mu} = -s^{2}_{24} s^{2}_{25} \sin 2\alpha^{}_2 \; ,
\nonumber \\
& {\cal Z}^{4 5}_{\tau \tau} = -s^{2}_{34} s^{2}_{35} \sin 2\alpha^{}_3 \; ;
\hspace{1.8cm}
\nonumber \\
& {\cal Z}^{4 6}_{e e} = +s^{2}_{14} s^{2}_{16} \sin 2\gamma^{}_1 \; ,
\nonumber \\
& {\cal Z}^{4 6}_{\mu \mu} = +s^{2}_{24} s^{2}_{26} \sin 2\gamma^{}_2 \; ,
\nonumber \\
& {\cal Z}^{4 6}_{\tau \tau} = +s^{2}_{34} s^{2}_{36} \sin 2\gamma^{}_3 \; ;
\hspace{1.8cm}
\nonumber \\
& {\cal Z}^{5 6}_{e e} = -s^{2}_{15} s^{2}_{16} \sin 2\beta^{}_1 \; ,
\nonumber \\
& {\cal Z}^{5 6}_{\mu \mu} = -s^{2}_{25} s^{2}_{26} \sin 2\beta^{}_2 \; ,
\nonumber \\
& {\cal Z}^{5 6}_{\tau \tau} = -s^{2}_{35} s^{2}_{36} \sin 2\beta^{}_3 \; ,
\hspace{1.8cm}
\label{16}
\end{align}
where the relevant CP-violating phases have been defined in Eq.~(\ref{14}).
It becomes clear that the phase combinations of
${\cal Z}^{j j^\prime}_{\alpha \beta}$ are quite different from those of
${\cal X}^{j j^\prime}_{\alpha \beta}$, although they both depend on the
same set of original seesaw phase parameters (i.e., $\alpha^{}_i$ and
$\beta^{}_i$ with $\gamma^{}_i = -\left(\alpha^{}_i + \beta^{}_i\right)$
for $i = 1, 2, 3$).

\subsection{The expressions of $\varepsilon^{}_{j \alpha}$}

Now that $R$ measures the strengths of weak charged-current interactions
associated with the three heavy Majorana neutrinos $N^{}_j$ (for $j = 4, 5, 6$)
in the canonical seesaw framework, its rephasing invariants listed above
must determine the CP-violating effects in the LNV decays
of $N^{}_j$. As the masses $M^{}_j$ of $N^{}_j$ are anticipated to be far above
the electroweak symmetry breaking scale characterized by the vacuum expectation
value of the Higgs field $\langle H\rangle \simeq 174~{\rm GeV}$,
one may simply calculate the CP-violating asymmetries between $N^{}_j$ decays
into the leptonic doublet $\ell^{}_\alpha$ plus the Higgs doublet $H$ and
their CP-conjugated processes (for $\alpha = e, \mu, \tau$) at the one-loop
level~\cite{Luty:1992un,Covi:1996wh,Plumacher:1996kc,Pilaftsis:1997jf,
Buchmuller:2005eh,DiBari:2021fhs}. With the help of our Euler-like
block parametrization of the seesaw flavor structure, we have the
flavor-dependent asymmetries~\cite{Xing:2024xwb}
\begin{eqnarray}
\varepsilon^{}_{j \alpha} \hspace{-0.2cm} & \equiv & \hspace{-0.2cm}
\frac{\Gamma({N}^{}_j \to \ell^{}_\alpha + H)
- \Gamma({N}^{}_j \to \overline{\ell^{}_\alpha} +
\overline{H})}{\displaystyle \sum_\alpha \left[\Gamma({N}^{}_j \to
\ell^{}_\alpha + H) + \Gamma({N}^{}_j \to \overline{\ell^{}_\alpha}
+ \overline{H})\right]}
\nonumber \\
\hspace{-0.2cm} & \simeq & \hspace{-0.2cm}
\frac{1}{\displaystyle 8\pi \langle H\rangle^2 D^{}_j}
\sum^6_{j^\prime =4} \Bigg\{ M^2_{j^\prime} \hspace{0.1cm} {\rm Im} \Bigg[
\left(R^*_{\alpha j} R^{}_{\alpha j^\prime}\right) \sum_\beta \Big[ \left(R^*_{\beta j}
R^{}_{\beta j^\prime}\right) \xi(x^{}_{j^\prime j}) + \left(R^{}_{\beta j}
R^*_{\beta j^\prime}\right) \zeta(x^{}_{j^\prime j})\Big] \Bigg] \Bigg\} \; ,
\hspace{0.5cm}
\label{17}
\end{eqnarray}
where the Latin and Greek subscripts run respectively over $(4, 5, 6)$ and
$(e, \mu, \tau)$, the two loop functions
$\xi(x^{}_{j^\prime j}) = \sqrt{x^{}_{j^\prime j}} \left\{1 + 1/\left(1 -
x^{}_{j^\prime j}\right) + \left(1 + x^{}_{j^\prime j}\right) \ln
\left[x^{}_{j^\prime j} / \left(1 + x^{}_{j^\prime j}\right)
\right] \right\}$ and $\zeta(x^{}_{j^\prime j}) = 1/\left(1 -
x^{}_{j^\prime j}\right)$ depend on the ratios $x^{}_{j^\prime j} \equiv
{M}^2_{j^\prime}/{M}^2_j$, and the factor $D^{}_j$ is defined as
$D^{}_j \equiv \left|R^{}_{e j}\right|^2 + \left|R^{}_{\mu j}\right|^2 +
\left|R^{}_{\tau j}\right|^2$.

Given the two sets of rephasing invariants
defined in Eq.~(\ref{5}), it is straightforward for us to rewrite nine
$\varepsilon^{}_{j \alpha}$ and their three flavor-independent counterparts
$\varepsilon^{}_j \equiv \varepsilon^{}_{j e} + \varepsilon^{}_{j \mu} +
\varepsilon^{}_{j \tau}$ as
\begin{eqnarray}
\varepsilon^{}_{j \alpha} \simeq
\frac{-1}{\displaystyle 8\pi \langle H\rangle^2 D^{}_j}
\sum^6_{j^\prime =4} \Bigg[ M^2_{j^\prime} \sum_\beta \Big[
{\cal Z}^{j j^\prime}_{\alpha \beta} \xi(x^{}_{j^\prime j}) +
{\cal X}^{j j^\prime}_{\alpha \beta} \zeta(x^{}_{j^\prime j})\Big] \Bigg] \; ,
\label{18}
\end{eqnarray}
and
\begin{eqnarray}
\varepsilon^{}_{j} \simeq
\frac{-1}{\displaystyle 8\pi \langle H\rangle^2 D^{}_j}
\sum^6_{j^\prime =4} \left[ M^2_{j^\prime} \left( \sum_\alpha
{\cal Z}^{j j^\prime}_{\alpha \alpha} + 2 \sum_{\alpha < \beta}
{\cal Z}^{j j^\prime}_{\alpha \beta} \right) \xi(x^{}_{j^\prime j}) \right] \; .
\label{19}
\end{eqnarray}
It is obvious that $\varepsilon^{}_j$ is insensitive to the rephasing
invariants ${\cal X}^{j j^\prime}_{\alpha \beta}$, and hence contains less
information on CP violation than $\varepsilon^{}_{j \alpha}$. As the
CP-violating asymmetries $\varepsilon^{}_{j \alpha}$ (or $\varepsilon^{}_{j}$)
play a central role in thermal leptogenesis, the application of
${\cal X}^{j j^\prime}_{\alpha \beta}$ and ${\cal Z}^{j j^\prime}_{\alpha \beta}$
to the description of $\varepsilon^{}_{j \alpha}$ and $\varepsilon^{}_{j}$
implies the usefulness of such rephasing invariants of CP violation in the
study of the cosmic matter-antimatter asymmetry issue with the help of the seesaw
and leptogenesis mechanisms.

\section{Non-unitarity effects in ${\cal J}^{i i^\prime}_{\alpha \beta}$}

In Ref.~\cite{Xing:2020ivm}, the expressions of ${\cal J}^{i i^\prime}_{\alpha \beta}$
were derived up to the accuracy of ${\cal O}(a^{}_{ii})$ and
${\cal O}(s^{2}_{13} \left|a^{}_{ii^\prime}\right|)$, where
$a^{}_{ii}$ and $a^{}_{ii^\prime}$ have been defined in Eq.~(\ref{12})
(for $ii = 11, 22, 33$ and $i i^\prime = 21, 31, 32$). Given the fact that
the Jarlskog invariant ${\cal J}^{}_\nu$ is proportional to $s^{}_{13}$
with $\theta^{}_{13}$ being the smallest active flavor mixing angle of
$U^{}_0$~\cite{DayaBay:2012fng,Capozzi:2025wyn}, we find that it will be
good enough to consider the deviations of
${\cal J}^{i i^\prime}_{\alpha \beta}$ from $\pm {\cal J}^{}_\nu$
to the degree of accuracy of ${\cal O}(a^{}_{ii})$ and
${\cal O}(s^{}_{13} \left|a^{}_{ii^\prime}\right|)$. Needless to say, such
deviations originate from the small non-unitarity of $U = A U^{}_0$
as compared with $U^{}_0$ in the seesaw framework.

Taking account of the approximations made in Eq.~(\ref{11}), we are left
with the PMNS matrix of the form
\begin{eqnarray}
U = U^{}_0 - \left( \begin{matrix} a^{}_{11} & 0 & 0 \cr
a^{}_{21} & a^{}_{22} & 0 \cr
a^{}_{31} & a^{}_{32} & a^{}_{33} \cr \end{matrix}
\right) U^{}_0 \; ,
\label{20}
\end{eqnarray}
where the magnitudes of $a^{}_{ii}$ and $a^{}_{ii^\prime}$ should be
at most of ${\cal O}(10^{-3})$, as indicated by the upper bounds of the
nine active-sterile flavor mixing angles in Eq.~(\ref{9}) or Eq.~(\ref{10}).
As ${\cal J}^{i i^\prime}_{\alpha \alpha} = {\cal J}^{i i}_{\alpha \beta} = 0$
and ${\cal J}^{i i^\prime}_{\beta \alpha} =
{\cal J}^{i^\prime i}_{\alpha \beta} = -{\cal J}^{i i^\prime}_{\alpha \beta}$
hold, we only need to calculate those of ${\cal J}^{i i^\prime}_{\alpha \beta}$
with $\alpha < \beta$ and $i < i^\prime$. The results are
\begin{align}
& {\cal J}^{12}_{e\mu} - {\cal J}^{}_\nu = - 2 \left(a^{}_{11} + a^{}_{22}\right)
{\cal J}^{}_\nu + c^{}_{12} s^{}_{12} c^{}_{13} c^{}_{23}
{\rm Im}\left(a^{}_{21} e^{-{\rm i}\delta^{}_{12}}\right) \; ,
\nonumber \\
& {\cal J}^{12}_{e\tau} + {\cal J}^{}_\nu = + 2 \left(a^{}_{11} + a^{}_{33}\right)
{\cal J}^{}_\nu - c^{}_{12} s^{}_{12} c^{}_{13} s^{}_{23}
{\rm Im}\left[a^{}_{31} e^{{\rm i} \left(\delta^{}_{13} - \delta^{}_\nu\right)}\right]
+ c^{}_{12} s^{}_{12} s^{}_{13} \left\{c^2_{23} {\rm Im}
\left[a^{}_{32} e^{{\rm i}\left(\delta^{}_{12} - \delta^{}_{13}\right)}\right]
\right.
\nonumber \\
& \left. \hspace{2.1cm}
+ \hspace{0.05cm}
s^2_{23} {\rm Im}\left[a^{}_{32} e^{{\rm i}\left(\delta^{}_\nu - \delta^{}_{23}\right)}
\right] \right\} \; ,  \hspace{0.31cm}
\nonumber \\
& {\cal J}^{12}_{\mu\tau} - {\cal J}^{}_\nu = - 2 \left(a^{}_{22} + a^{}_{33}\right)
{\cal J}^{}_\nu + c^{}_{13} c^{}_{23} s^{}_{23} \left\{c^{}_{12} s^{}_{12} s^{}_{23}
{\rm Im}\left(a^{}_{21} e^{-{\rm i}\delta^{}_{12}}\right)
+ \left(c^2_{12} - s^2_{12}\right) s^{}_{13} c^{}_{23}
{\rm Im}\left[a^{}_{21} e^{{\rm i}\left(\delta^{}_{23} - \delta^{}_{13}\right)}\right]
\right\}
\nonumber \\
& \hspace{2.15cm}
+ c^{}_{13} c^{}_{23} s^{}_{23} \left\{c^{}_{12} s^{}_{12} c^{}_{23}
{\rm Im}\left[a^{}_{31} e^{{\rm i} \left(\delta^{}_\nu - \delta^{}_{13}\right)}\right]
+ \left(c^2_{12} - s^2_{12}\right)
s^{}_{13} s^{}_{23} {\rm Im}\left(a^{}_{31} e^{-{\rm i}\delta^{}_{13}}\right) \right\}
\nonumber \\
& \hspace{2.15cm}
- c^{}_{12} s^{}_{12} s^{}_{13} c^{2}_{23} \left\{ 2 s^{2}_{23} \cos\delta^{}_\nu
\hspace{0.05cm} {\rm Im}\left(a^{}_{32} e^{-{\rm i}\delta^{}_{23}}\right)
+ c^2_{23} {\rm Im}\left[a^{}_{32} e^{{\rm i}\left(\delta^{}_{12} - \delta^{}_{13}\right)}
\right] - s^2_{23} {\rm Im}\left[ a^{}_{32}
e^{{\rm i}\left(\delta^{}_\nu - \delta^{}_{23}\right)}
\right]\right\} \; ;
\nonumber \\
& {\cal J}^{13}_{e\mu} + {\cal J}^{}_\nu
= + 2 \left(a^{}_{11} + a^{}_{22}\right) {\cal J}^{}_\nu
+ c^{2}_{12} c^{}_{13} s^{}_{13} s^{}_{23}
{\rm Im}\left[a^{}_{21} e^{{\rm i} \left(\delta^{}_{23} - \delta^{}_{13}\right)}\right] \; ,
\nonumber \\
& {\cal J}^{13}_{e\tau} - {\cal J}^{}_\nu
= - 2 \left(a^{}_{11} + a^{}_{33}\right) {\cal J}^{}_\nu
+ c^{2}_{12} c^{}_{13} s^{}_{13} c^{}_{23}
{\rm Im}\left(a^{}_{31} e^{-{\rm i}\delta^{}_{13}}\right)
- c^{}_{12} s^{}_{12} s^{}_{13} \left\{ c^2_{23}
{\rm Im}\left[a^{}_{32} e^{{\rm i}\left(\delta^{}_{12} - \delta^{}_{13}\right)}\right]
\right.
\nonumber \\
& \left. \hspace{2.1cm}
+ \hspace{0.05cm} s^2_{23} {\rm Im}\left[a^{}_{32} e^{{\rm i}\left(\delta^{}_\nu -
\delta^{}_{23}\right)}\right]\right\} \; ,
\nonumber \\
& {\cal J}^{13}_{\mu\tau} + {\cal J}^{}_\nu
= + 2 \left(a^{}_{22} + a^{}_{33}\right) {\cal J}^{}_\nu
- c^{}_{13} c^{}_{23} s^{}_{23} \left\{c^{}_{12} s^{}_{12} s^{}_{23}
{\rm Im}\left(a^{}_{21} e^{-{\rm i}\delta^{}_{12}}\right)
- \left(c^2_{12} - s^2_{12}\right) s^{}_{13} c^{}_{23}
{\rm Im}\left[a^{}_{21} e^{{\rm i}\left(\delta^{}_{23} - \delta^{}_{13}\right)}
\right]\right\}
\nonumber \\
& \hspace{2.15cm}
- c^{}_{13} c^{}_{23} s^{}_{23} \left\{c^{}_{12} s^{}_{12} c^{}_{23}
{\rm Im}\left[a^{}_{31} e^{{\rm i}\left(\delta^{}_\nu - \delta^{}_{13}\right)}\right]
+ \left(c^2_{12} - s^2_{12}\right) s^{}_{13}
s^{}_{23} {\rm Im}\left(a^{}_{31} e^{-{\rm i}\delta^{}_{13}}\right) \right\}
\nonumber \\
& \hspace{2.15cm}
+ s^{}_{12} c^{}_{23} s^{}_{23}
\left(s^{}_{12} + 2 c^{}_{12} s^{}_{13} c^{}_{23} s^{}_{23} \cos\delta^{}_\nu\right)
{\rm Im}\left(a^{}_{32} e^{-{\rm i}\delta^{}_{23}}\right)
\nonumber \\
& \hspace{2.15cm} - c^{}_{12} s^{}_{12} s^{}_{13} s^2_{23}
\left\{c^2_{23} {\rm Im}\left[a^{}_{32}
e^{{\rm i}\left(\delta^{}_{12} - \delta^{}_{13}\right)}\right]
- s^2_{23} {\rm Im}\left[a^{}_{32}
e^{{\rm i}\left(\delta^{}_\nu - \delta^{}_{23}\right)}\right] \right\} \; ;
\nonumber \\
& {\cal J}^{23}_{e\mu} - {\cal J}^{}_\nu
= - 2 \left(a^{}_{11} + a^{}_{22}\right) {\cal J}^{}_\nu
+ s^2_{12} c^{}_{13} s^{}_{13} s^{}_{23} {\rm Im}\left[a^{}_{21}
e^{{\rm i}\left(\delta^{}_{23} - \delta^{}_{13}\right)}\right] \; ,
\nonumber \\
& {\cal J}^{23}_{e\tau} + {\cal J}^{}_\nu
= + 2 \left(a^{}_{11} + a^{}_{33}\right) {\cal J}^{}_\nu
+ s^{2}_{12} c^{}_{13} s^{}_{13} c^{}_{23} {\rm Im}\left(a^{}_{31}
e^{-{\rm i}\delta^{}_{13}}\right)
+ c^{}_{12} s^{}_{12} s^{}_{13} \left\{ c^2_{23}
{\rm Im}\left[a^{}_{32} e^{{\rm i}\left(\delta^{}_{12} - \delta^{}_{13}\right)}\right]
\right.
\nonumber \\
& \left. \hspace{2.1cm}
+ \hspace{0.05cm} s^2_{23} {\rm Im}\left[a^{}_{32} e^{{\rm i}
\left(\delta^{}_\nu - \delta^{}_{23}\right)}\right]\right\} \; ,
\nonumber \\
& {\cal J}^{23}_{\mu\tau} - {\cal J}^{}_\nu
= - 2 \left(a^{}_{22} + a^{}_{33}\right) {\cal J}^{}_\nu
+ c^{}_{13} c^{}_{23} s^{}_{23} \left\{c^{}_{12} s^{}_{12} s^{}_{23}
{\rm Im}\left(a^{}_{21} e^{-{\rm i}\delta^{}_{12}}\right)
- \left(c^2_{12} - s^2_{12}\right) s^{}_{13} c^{}_{23}
{\rm Im}\left[a^{}_{21} e^{{\rm i}\left(\delta^{}_{23} - \delta^{}_{13}\right)}
\right]\right\}
\nonumber \\
& \hspace{2.15cm}
+ c^{}_{13} c^{}_{23} s^{}_{23} \left\{ c^{}_{12} s^{}_{12} c^{}_{23}
{\rm Im}\left[a^{}_{31} e^{{\rm i}\left(\delta^{}_\nu - \delta^{}_{13}\right)}\right]
+ \left(c^2_{12} - s^2_{12}\right) s^{}_{13} s^{}_{23}
{\rm Im}\left(a^{}_{31} e^{-{\rm i}\delta^{}_{13}}\right)\right\}
\nonumber \\
& \hspace{2.15cm}
+ c^{}_{12} c^{}_{23} s^{}_{23} \left(c^{}_{12} - 2 s^{}_{12} s^{}_{13} c^{}_{23}
s^{}_{23} \cos\delta^{}_\nu\right) {\rm Im}\left(a^{}_{32}
e^{-{\rm i}\delta^{}_{23}}\right)
\nonumber \\
& \hspace{2.15cm}
+ c^{}_{12} s^{}_{12} s^{}_{13} s^2_{23}
\left\{ c^2_{23} {\rm Im}\left[a^{}_{32} e^{{\rm i}\left(\delta^{}_{12}
- \delta^{}_{13}\right)}\right] - s^2_{23} {\rm Im}\left[a^{}_{32}
e^{{\rm i}\left(\delta^{}_\nu - \delta^{}_{23}\right)}\right]\right\} \; ,
\label{21}
\end{align}
where
\begin{eqnarray}
{\cal J}^{}_\nu = c^{}_{12} s^{}_{12} c^2_{13} s^{}_{13} c^{}_{23} s^{}_{23}
\sin\delta^{}_\nu \;
\label{22}
\end{eqnarray}
is just the universal Jarlskog invariant of CP violation in the unitarity limit of
$U$, and $\delta^{}_\nu \equiv \delta^{}_{13} - \delta^{}_{12} - \delta^{}_{23}$
has been defined. It becomes transparent that the leading terms of all the nontrivial
${\cal J}^{i i^\prime}_{\alpha \beta}$ are universally $\pm {\cal J}^{}_\nu$.
Unless the magnitude of ${\cal J}^{}_\nu$ were highly suppressed due
to $\delta^{}_\nu \to 0$ or $\pi$, the non-unitarity contributions to
${\cal J}^{i i^\prime}_{\alpha \beta}$ would not play an important role
due to the suppression of active-sterile flavor mixing.
Nevertheless, the future long-baseline precision measurements might be able
to constrain some of the non-unitarity effects hidden in
${\cal J}^{i i^\prime}_{\alpha \beta}$~\cite{DUNE:2015lol,Hyper-KamiokandeProto-:2015xww}.

To see this point clearly, one may write out the probabilities of
$\nu^{}_\alpha \to \nu^{}_\beta$ and
$\overline{\nu}^{}_\alpha \to \overline{\nu}^{}_\beta$
oscillations~\cite{Xing:2007zj,Xing:2020ivm}:
\begin{eqnarray}
&& P(\nu^{}_\alpha \to \nu^{}_\beta) = \frac{1}{D^{}_{\alpha\beta}}
\left[ \sum^{3}_{i=1} |U^{}_{\alpha i}|^2 |U^{}_{\beta i}|^2 + 2 \sum^{}_{i<i^\prime}
{\rm Re} \left( U^{}_{\alpha i} U^{}_{\beta i^\prime} U^*_{\alpha i^\prime}
U^*_{\beta i} \right) \cos \Delta^{}_{i^\prime i} + 2 \sum^{}_{i<i^\prime}
{\cal J}^{ii^\prime}_{\alpha\beta} \sin\Delta^{}_{i^\prime i}\right] \; ,
\nonumber \\
&& P(\overline{\nu}^{}_\alpha \to \overline{\nu}^{}_\beta) =
\frac{1}{D^{}_{\alpha\beta}}
\left[ \sum^{3}_{i=1} |U^{}_{\alpha i}|^2 |U^{}_{\beta i}|^2 + 2 \sum^{}_{i<i^\prime}
{\rm Re} \left( U^{}_{\alpha i} U^{}_{\beta i^\prime} U^*_{\alpha i^\prime}
U^*_{\beta i} \right) \cos \Delta^{}_{i^\prime i} - 2 \sum^{}_{i<i^\prime}
{\cal J}^{ii^\prime}_{\alpha\beta} \sin\Delta^{}_{i^\prime i}\right] \; , \hspace{1cm}
\label{23}
\end{eqnarray}
where $D^{}_{\alpha\beta} \equiv \left(UU^\dagger\right)^{}_{\alpha\alpha}
\left(UU^\dagger\right)^{}_{\beta\beta} = \left(AA^\dagger\right)^{}_{\alpha\alpha}
\left(AA^\dagger\right)^{}_{\beta\beta}$,
$\Delta^{}_{i^\prime i} \equiv \Delta m^2_{i^\prime i} L/\left(2E\right)$ with $\Delta
m^2_{i^\prime i} \equiv m^2_{i^\prime} - m^2_i$ (for $i^\prime i = 21, 31, 32$),
$E$ being the average neutrino beam energy and $L$ being the baseline length.
Therefore, we arrive at the CP-violating asymmetries
\begin{eqnarray}
{\cal A}^{}_{\alpha\beta} \equiv
P(\nu^{}_\alpha \to \nu^{}_\beta) - P(\overline{\nu}^{}_\alpha \to
\overline{\nu}^{}_\beta) = \frac{4}{D^{}_{\alpha\beta}}
\sum^{}_{i<i^\prime} {\cal J}^{ii^\prime}_{\alpha\beta} \sin\Delta^{}_{i^\prime i} \; .
\label{24}
\end{eqnarray}
To be more explicit, let us consider $\nu^{}_\mu \to \nu^{}_e$ and
$\overline{\nu}^{}_\mu \to \overline{\nu}^{}_e$ oscillations for
example. In this most interesting case the CP-violating asymmetry
${\cal A}^{}_{\mu e}$ turns out to be
\begin{eqnarray}
{\cal A}^{}_{\mu e} \hspace{-0.2cm} & \simeq & \hspace{-0.2cm}
-16 {\cal J}^{}_\nu \sin\frac{\Delta^{}_{21}}{2}
\sin\frac{\Delta^{}_{31}}{2} \sin\frac{\Delta^{}_{32}}{2}
- 4 c^{}_{13} \Big\{ c^{}_{12} s^{}_{12} c^{}_{23}
{\rm Im}\left(a^{}_{21} e^{-{\rm i}\delta^{}_{21}}\right) \sin\Delta^{}_{21}
\nonumber \\
\hspace{-0.2cm} & & \hspace{-0.2cm}
+ \hspace{0.05cm} c^2_{12} s^{}_{13} s^{}_{23} {\rm Im}\left[a^{}_{21} e^{{\rm i}
\left(\delta^{}_{23} - \delta^{}_{13}\right)}\right] \sin\Delta^{}_{31}
+ s^2_{12} s^{}_{13} s^{}_{23} {\rm Im}\left[a^{}_{21} e^{{\rm i}
\left(\delta^{}_{23} - \delta^{}_{13}\right)}\right] \sin\Delta^{}_{32}
\Big\} \; , \hspace{1cm}
\label{25}
\end{eqnarray}
where $D^{}_{\mu e} \simeq 1 - 2\left(a^{}_{11} + a^{}_{22}\right)$ has been
taken into account to cancel the coefficient of ${\cal J}^{}_\nu$ in the
expressions of ${\cal J}^{12}_{e\mu}$, ${\cal J}^{13}_{e\mu}$ and
${\cal J}^{23}_{e\mu}$ obtained in Eq.~(\ref{21}), and
the terms associated with $a^{}_{21}$ characterize the contributions
induced by non-unitarity of the PMNS matrix.
In practice, terrestrial matter effects on neutrino oscillations have to
be taken into consideration for a realistic long-baseline experiment
\footnote{Terrestrial matter effects on T violation are less significant
in a long-baseline neutrino oscillation experiments~\cite{Xing:2013uxa,
Schwetz:2021cuj}, but establishing a genuine
signal of T violation is practically challenging although it is
characterized by the same ${\cal J}^{}_\nu$.},
and they are unavoidably entangled with the small non-unitarity effects
originating from nonzero $R$ in the seesaw mechanism (see, e.g.,
Refs.~\cite{Li:2015oal,Xing:2016ymg,Li:2018jgd} for some detailed discussions).

\section{Comments on $\mathbb{X}^{i j}_{\alpha \beta}$ and
$\mathbb{Z}^{i j}_{\alpha \beta}$}

As for some LFV and LNV processes mediated by both the light Majorana neutrino
fields $\nu^{}_i$ (for $i = 1, 2, 3$) and the heavy Majorana neutrino fields
$N^{}_j$ (for $j = 4, 5, 6$) in the canonical seesaw mechanism, the interplay
between the contributions associated with the PMNS matrix $U$ and its counterpart
$R$ is expected to emerge, although one of them is likely to be strongly suppressed
as compared with the other. Let us illustrate such interplay effects, which are
characterized by the rephasing invariants $\mathbb{X}^{i j}_{\alpha \beta}$ and
$\mathbb{Z}^{i j}_{\alpha \beta}$ defined in Eq.~(\ref{6}), by taking two
typical examples as follows.

{\bf Example (1):} the one-loop radiative decays of charged leptons of the form
$\alpha^- \to \beta^- + \gamma$ (for $\alpha^- = \mu^-$ or $\tau^-$, and
$\beta^- = e^-$ or $\mu^-$) with $\gamma$ being the photon, which belong to
the LFV processes. In the seesaw framework we find that their branching ratios
can be expressed as
\begin{eqnarray}
{\cal B}^{}_{\alpha \to \beta + \gamma} \hspace{-0.2cm} & = & \hspace{-0.2cm}
\frac{3 \alpha^{}_{\rm em}}{2 \pi} \left|\sum^3_{i=1} U^*_{\alpha i}
U^{}_{\beta i} G^{}_\gamma(x^{}_i) + \sum^6_{j=4} R^*_{\alpha j} R^{}_{\beta j}
G^{}_\gamma(x^{\prime}_j)\right|^2 \;
\nonumber \\
\hspace{-0.2cm} & = & \hspace{-0.2cm}
\frac{3 \alpha^{}_{\rm em}} {32 \pi} \left[
\left|\sum^3_{i=1} U^*_{\alpha i} U^{}_{\beta i} G^{}_\gamma(x^{}_i)\right|^2
+ 2 \sum^3_{i=1} \sum^6_{j=4} {\rm Re} \mathbb{X}^{i j}_{\alpha \beta}
G^{}_\gamma(x^{}_i) G^{}_\gamma(x^{\prime}_j)
+ \left|\sum^6_{j=4} R^*_{\alpha j} R^{}_{\beta j}
G^{}_\gamma(x^{\prime}_j)\right|^2 \right] \; , \hspace{0.5cm}
\label{26}
\end{eqnarray}
where $\alpha^{}_{\rm em}$ is the fine structure constant of electromagnetic
interactions, $x^{}_i \equiv m^2_i/M^2_W$ and $x^\prime_j \equiv M^2_j/M^2_W$
(for $i = 1,2,3$ and $j = 4, 5, 6$), $G^{}_\gamma (x^{}_i)$ and
$G^{}_\gamma (x^{\prime}_j)$ stand respectively
for the loop functions of light and heavy
Majorana neutrinos~\cite{Ilakovac:1994kj,Alonso:2012ji,Xing:2020ijf}. We see
that the rephasing invariants ${\rm Re} \mathbb{X}^{i j}_{\alpha \beta}$
have appeared in ${\cal B}^{}_{\alpha \to \beta + \gamma}$, but they are
apparently CP-conserving.

{\bf Example (2):} the $0\nu 2\beta$ decays mediated by both $\nu^{}_i$
and $N^{}_j$, whose overall width can be expressed as
\begin{eqnarray}
\Gamma^{}_{0\nu 2\beta} \hspace{-0.2cm} & \propto & \hspace{-0.2cm}
\left|\sum^3_{i=1} m^{}_i U^2_{e i} - M^2_A\sum^6_{j=4}
\frac{R^2_{e j}}{M^{}_j} {\cal F}(A, M^{}_j)\right|^2 \;
\nonumber \\
\hspace{-0.2cm} & \propto & \hspace{-0.2cm}
\left[\left|\sum^3_{i=1} m^{}_i U^2_{e i}\right|^2 - 2 M^2_A
\sum^3_{i=1} \sum^6_{j=4} \frac{m^{}_i}{M^{}_j} {\rm Re} \mathbb{Z}^{i j}_{ee}
{\cal F}(A, M^{}_j)
+ M^4_A \left|\sum^6_{j=4} \frac{R^2_{e j}}{M^{}_j} {\cal F}(A, M^{}_j)\right|^2
\right] \; , \hspace{0.5cm}
\label{27}
\end{eqnarray}
where $A$ denotes the atomic number of a given isotope, ${\cal F}(A, M^{}_i) \simeq 0.1$
depends mildly on the decaying nucleus, and $M^{}_A \simeq 0.1$ GeV
\cite{Haxton:1984ggj,Fang:2021jfv,Fang:2024hzy}. It is obvious that the rephasing
invariants ${\rm Re} \mathbb{Z}^{i j}_{ee}$ have shown up in $\Gamma^{}_{0\nu 2\beta}$,
and they are also CP-conserving
\footnote{Note, however, that the light and heavy Majorana neutrino terms in
Eq.~(\ref{27}) are intrinsically correlated with each other via the seesaw
relation given by Eq.~(\ref{2}), namely
$(U D^{}_\nu U^T)^{}_{ee} = -(R D^{}_N R^T)^{}_{ee}$~\cite{Xing:2009ce}.
In this case Eq.~(\ref{27}) can be simplified to
$\displaystyle \Gamma^{}_{0\nu 2\beta} \propto
\left|\sum^6_{i=4} \left[M^{}_j - \frac{M^2_A}{M^{}_j} {\cal F}(A, M^{}_j)
\right] R^2_{e j} \right|^2$.}.
Similar discussions can be extended to the hadronic LNV processes like
$B^-_u \to \pi^+ \alpha^- \beta^-$ (for $\alpha, \beta = e, \mu, \tau$),
although it is more difficult to observe such rare decay
modes~\cite{Doi:1985dx,Xing:2022hst} in a high-energy accelerator experiment.

In summary, we have written out the Majorana-type invariants of CP violation
of $U$ and $R$ in the canonical seesaw mechanism (i.e.,
${\cal V}^{ii^\prime}_{\alpha\beta}$ and ${\cal Z}^{j j^\prime}_{\alpha\beta}$
for $i, i^\prime = 1, 2, 3$ and $j, j^\prime = 4, 5, 6$, together with
$\alpha, \beta, \gamma = e, \mu, \tau$), which are insensitive to rephasing
three charged-lepton fields; and the Dirac-type invariants of CP violation
of $U$ and $R$ (i.e., ${\cal J}^{ii^\prime}_{\alpha\beta}$ and
${\cal X}^{j j^\prime}_{\alpha\beta}$) that are insensitive
to the rephasing of both the charged-lepton fields and the neutrino fields.
With the help of a full Euler-like block parametrization of the seesaw flavor
structure containing nine active-sterile flavor mixing angles and six
independent CP-violating phases, we have explicitly calculated
${\cal X}^{j j^\prime}_{\alpha\beta}$ and ${\cal Z}^{j j^\prime}_{\alpha\beta}$,
and briefly discussed their roles in the CP-violating asymmetries of three heavy
Majorana neutrino decays within the thermal leptogenesis framework.
The non-unitarity-induced deviations of
${\cal J}^{ii^\prime}_{\alpha\beta}$ from the universal Jarlskog invariant
$\pm {\cal J}^{}_\nu$ have been examined, and their small effects in
the flavor oscillations of three light Majorana neutrinos have been briefly
discussed. We have also pointed out that the rephasing invariants
$\mathbb{X}^{i j}_{\alpha \beta}$ and $\mathbb{Z}^{i j}_{\alpha \beta}$,
which originate from the interplay between $R$ and $U$, may manifest themselves
in some interesting LFV and LNV processes.

We believe that the rephasing invariants of leptonic CP violation proposed
and discussed in this work will find more applications in the studies of
neutrino phenomenology and neutrino cosmology, and they can be extended
to the inverse and linear seesaw mechanisms in a straightforward
way~\cite{Han:2021qum}. Different from the basis-independent flavor invariants
constructed from the seesaw-associated effective field theory (see, e.g.,
Refs.~\cite{Wang:2021wdq,Yu:2021cco,Yu:2022nxj,Yu:2022ttm}),
our rephasing invariants are constructed in the mass basis of six Majorana
neutrino fields and hence not only completely independent of the mass parameters
but also formally much simpler.

\section*{Acknowledgements}

The author is greatly indebted to Jihong Huang, Di Zhang, Shun Zhou and
Jing-yu Zhu for some useful discussions, and especially to Di and Jihong
for kindly pointing out the typos in my original manuscript. This research
was supported in part by the Scientific and Technological Innovation Program
of the Institute of High Energy Physics under Grant No. E55457U2.


\begin{thebibliography}{99}

\bibitem{Minkowski:1977sc}
P.~Minkowski,
``$\mu \to e\gamma$ at a rate of one out of $10^{9}$ muon decays?,''
Phys.\ Lett.\  {\bf 67B} (1977) 421.

\bibitem{Yanagida:1979as}
T.~Yanagida,
``Horizontal gauge symmetry and masses of neutrinos,''
Conf.\ Proc.\ C {\bf 7902131} (1979) 95.

\bibitem{GellMann:1980vs}
M.~Gell-Mann, P.~Ramond and R.~Slansky,
``Complex spinors and unified theories,''
Conf.\ Proc.\ C {\bf 790927} (1979) 315
[arXiv:1306.4669 [hep-th]].

\bibitem{Glashow:1979nm}
S.~L.~Glashow,
``The future of elementary particle physics,''
NATO Sci.\ Ser.\ B {\bf 61} (1980) 687.

\bibitem{Mohapatra:1979ia}
R.~N.~Mohapatra and G.~Senjanovic,
``Neutrino mass and spontaneous parity nonconservation,''
Phys.\ Rev.\ Lett.\  {\bf 44} (1980) 912.

\bibitem{Fukugita:1986hr}
M.~Fukugita and T.~Yanagida,
``Baryogenesis Without Grand Unification,''
Phys. Lett. B \textbf{174} (1986), 45-47.

\bibitem{Pontecorvo:1957cp}
B.~Pontecorvo,
``Mesonium and anti-mesonium,''
Sov.\ Phys.\ JETP {\bf 6} (1957) 429
[Zh.\ Eksp.\ Teor.\ Fiz.\  {\bf 33} (1957) 549].

\bibitem{Maki:1962mu}
Z.~Maki, M.~Nakagawa and S.~Sakata,
``Remarks on the unified model of elementary particles,''
Prog.\ Theor.\ Phys.\  {\bf 28} (1962) 870.

\bibitem{Pontecorvo:1967fh}
B.~Pontecorvo,
``Neutrino Experiments and the Problem of Conservation of Leptonic Charge,''
Sov.\ Phys.\ JETP {\bf 26} (1968) 984
[Zh.\ Eksp.\ Teor.\ Fiz.\  {\bf 53} (1967) 1717].

\bibitem{Xing:2007zj}
Z.~z.~Xing,
``Correlation between the Charged Current Interactions of Light and Heavy Majorana Neutrinos,''
Phys. Lett. B \textbf{660} (2008), 515-521
[arXiv:0709.2220 [hep-ph]].

\bibitem{Xing:2011ur}
Z.~z.~Xing,
``A full parametrization of the 6$\times$6 flavor mixing matrix in the presence of three light
or heavy sterile neutrinos,''
Phys. Rev. D \textbf{85} (2012), 013008
[arXiv:1110.0083 [hep-ph]].

\bibitem{Jarlskog:1985ht}
C.~Jarlskog,
``Commutator of the Quark Mass Matrices in the Standard Electroweak Model and a
Measure of Maximal CP Nonconservation,''
Phys. Rev. Lett. \textbf{55} (1985), 1039

\bibitem{Wu:1985ea}
D.~d.~Wu,
``The Rephasing Invariants and CP,''
Phys. Rev. D \textbf{33} (1986), 860

\bibitem{Barger:1980jm}
V.~D.~Barger, K.~Whisnant and R.~J.~N.~Phillips,
``$CP$~Nonconservation in Three-Neutrino Oscillations,''
Phys. Rev. Lett. \textbf{45} (1980), 2084

\bibitem{Cheng:1986in}
H.~Y.~Cheng,
``{Kobayashi-Maskawa} Type of Hard {CP} Violation Model With Three Generation Majorana Neutrinos,''
Phys. Rev. D \textbf{34} (1986), 2794

\bibitem{Antusch:2006vwa}
S.~Antusch, C.~Biggio, E.~Fernandez-Martinez, M.~B.~Gavela and J.~Lopez-Pavon,
``Unitarity of the Leptonic Mixing Matrix,''
JHEP \textbf{10} (2006), 084
[arXiv:hep-ph/0607020 [hep-ph]].

\bibitem{Blennow:2016jkn}
M.~Blennow, P.~Coloma, E.~Fernandez-Martinez, J.~Hernandez-Garcia and J.~Lopez-Pavon,
``Non-Unitarity, sterile neutrinos, and Non-Standard neutrino Interactions,''
JHEP \textbf{04} (2017), 153
[arXiv:1609.08637 [hep-ph]].

\bibitem{Blennow:2023mqx}
M.~Blennow, E.~Fern\'andez-Mart\'\i{}nez, J.~Hern\'andez-Garc\'\i{}a, J.~L\'opez-Pav\'on,
X.~Marcano and D.~Naredo-Tuero,
``Bounds on lepton non-unitarity and heavy neutrino mixing,''
JHEP \textbf{08} (2023), 030
[arXiv:2306.01040 [hep-ph]].

\bibitem{Xing:2024gmy}
Z.~z.~Xing and J.~y.~Zhu,
``Confronting the seesaw mechanism with neutrino oscillations: A general and explicit analytical bridge,''
Nucl. Phys. B \textbf{1018} (2025), 117041
[arXiv:2412.17698 [hep-ph]].

\bibitem{Luo:2011mm}
S.~Luo,
``Dirac Lepton Angle Matrix v.s. Majorana Lepton Angle Matrix and Their Renormalization Group Running Behaviours,''
Phys. Rev. D \textbf{85} (2012), 013006
[arXiv:1109.4260 [hep-ph]].

\bibitem{Xing:2013ty}
Z.~z.~Xing,
``Properties of CP Violation in Neutrino-Antineutrino Oscillations,''
Phys. Rev. D \textbf{87} (2013) no.5, 053019
[arXiv:1301.7654 [hep-ph]].

\bibitem{Xing:2013woa}
Z.~z.~Xing and Y.~L.~Zhou,
``Majorana CP-violating phases in neutrino-antineutrino oscillations and other lepton-number-violating processes,''
Phys. Rev. D \textbf{88} (2013), 033002
[arXiv:1305.5718 [hep-ph]].

\bibitem{Wang:2021rsi}
Y.~Wang and S.~Zhou,
``Non-unitary leptonic flavor mixing and CP violation in neutrino-antineutrino oscillations,''
Phys. Lett. B \textbf{824} (2022), 136797
[arXiv:2109.13622 [hep-ph]].

\bibitem{Xing:2024xwb}
Z.~z.~Xing,
``Mapping the sources of CP violation in neutrino oscillations from the seesaw mechanism,''
Phys. Lett. B \textbf{856} (2024), 138909
[arXiv:2406.01142 [hep-ph]].

\bibitem{Luty:1992un}
M.~A.~Luty,
``Baryogenesis via leptogenesis,''
Phys. Rev. D \textbf{45} (1992), 455-465.

\bibitem{Covi:1996wh}
L.~Covi, E.~Roulet and F.~Vissani,
``CP violating decays in leptogenesis scenarios,''
Phys. Lett. B \textbf{384} (1996), 169-174
[arXiv:hep-ph/9605319 [hep-ph]].

\bibitem{Plumacher:1996kc}
M.~Plumacher,
``Baryogenesis and lepton number violation,''
Z. Phys. C \textbf{74} (1997), 549-559
[arXiv:hep-ph/9604229 [hep-ph]].

\bibitem{Pilaftsis:1997jf}
A.~Pilaftsis,
``CP violation and baryogenesis due to heavy Majorana neutrinos,''
Phys. Rev. D \textbf{56} (1997), 5431-5451
[arXiv:hep-ph/9707235 [hep-ph]].

\bibitem{Buchmuller:2005eh}
W.~Buchmuller, R.~D.~Peccei and T.~Yanagida,
``Leptogenesis as the origin of matter,''
Ann. Rev. Nucl. Part. Sci. \textbf{55} (2005), 311-355
[arXiv:hep-ph/0502169 [hep-ph]].

\bibitem{DiBari:2021fhs}
P.~Di Bari,
``On the origin of matter in the Universe,''
Prog. Part. Nucl. Phys. \textbf{122} (2022), 103913
[arXiv:2107.13750 [hep-ph]].

\bibitem{Xing:2020ivm}
Z.~z.~Xing and D.~Zhang,
``Radiative decays of charged leptons as constraints of unitarity polygons for active-sterile
neutrino mixing and CP violation,''
Eur. Phys. J. C \textbf{80} (2020) no.12, 1134
[arXiv:2009.09717 [hep-ph]].

\bibitem{DayaBay:2012fng}
F.~P.~An \textit{et al.} [Daya Bay],
``Observation of electron-antineutrino disappearance at Daya Bay,''
Phys. Rev. Lett. \textbf{108} (2012), 171803
[arXiv:1203.1669 [hep-ex]].

\bibitem{Capozzi:2025wyn}
F.~Capozzi, W.~Giar{\`e}, E.~Lisi, A.~Marrone, A.~Melchiorri and A.~Palazzo,
``Neutrino masses and mixing: Entering the era of subpercent precision,''
Phys. Rev. D \textbf{111} (2025) no.9, 093006
[arXiv:2503.07752 [hep-ph]].

\bibitem{DUNE:2015lol}
R.~Acciarri \textit{et al.} [DUNE],
``Long-Baseline Neutrino Facility (LBNF) and Deep Underground Neutrino Experiment (DUNE):
Conceptual Design Report, Volume 2: The Physics Program for DUNE at LBNF,''
[arXiv:1512.06148 [physics.ins-det]].

\bibitem{Hyper-KamiokandeProto-:2015xww}
K.~Abe \textit{et al.} [Hyper-Kamiokande Proto-],
``Physics potential of a long-baseline neutrino oscillation experiment using a J-PARC
neutrino beam and Hyper-Kamiokande,''
PTEP \textbf{2015} (2015), 053C02
[arXiv:1502.05199 [hep-ex]].

\bibitem{Xing:2013uxa}
Z.~z.~Xing,
``Leptonic commutators and clean T violation in neutrino oscillations,''
Phys. Rev. D \textbf{88} (2013) no.1, 017301
[arXiv:1304.7606 [hep-ph]].

\bibitem{Schwetz:2021cuj}
T.~Schwetz and A.~Segarra,
``Model-Independent Test of T Violation in Neutrino Oscillations,''
Phys. Rev. Lett. \textbf{128} (2022) no.9, 091801
[arXiv:2106.16099 [hep-ph]].

\bibitem{Li:2015oal}
Y.~F.~Li and S.~Luo,
``Neutrino Oscillation Probabilities in Matter with Direct and Indirect Unitarity Violation
in the Lepton Mixing Matrix,''
Phys. Rev. D \textbf{93} (2016) no.3, 033008
[arXiv:1508.00052 [hep-ph]].

\bibitem{Xing:2016ymg}
Z.~z.~Xing and J.~y.~Zhu,
``Analytical approximations for matter effects on CP violation in the accelerator-based neutrino
oscillations with E $\lesssim$ 1 GeV,''
JHEP \textbf{07} (2016), 011
[arXiv:1603.02002 [hep-ph]].

\bibitem{Li:2018jgd}
Y.~F.~Li, Z.~z.~Xing and J.~y.~Zhu,
``Indirect unitarity violation entangled with matter effects in reactor antineutrino oscillations,''
Phys. Lett. B \textbf{782} (2018), 578-588
[arXiv:1802.04964 [hep-ph]].

\bibitem{Ilakovac:1994kj}
A.~Ilakovac and A.~Pilaftsis,
``Flavor violating charged lepton decays in seesaw-type models,''
Nucl. Phys. B \textbf{437} (1995), 491
[arXiv:hep-ph/9403398 [hep-ph]].

\bibitem{Alonso:2012ji}
R.~Alonso, M.~Dhen, M.~B.~Gavela and T.~Hambye,
``Muon conversion to electron in nuclei in type-I seesaw models,''
JHEP \textbf{01} (2013), 118
[arXiv:1209.2679 [hep-ph]].

\bibitem{Xing:2020ijf}
Z.~z.~Xing,
``Flavor structures of charged fermions and massive neutrinos,''
Phys. Rept. \textbf{854} (2020), 1-147
[arXiv:1909.09610 [hep-ph]].

\bibitem{Haxton:1984ggj}
W.~C.~Haxton and G.~J.~Stephenson,
``Double beta Decay,''
Prog. Part. Nucl. Phys. \textbf{12} (1984), 409-479

\bibitem{Fang:2021jfv}
D.~L.~Fang, Y.~F.~Li and Y.~Y.~Zhang,
``Neutrinoless double beta decay in the minimal type-I seesaw model: How the enhancement or cancellation happens?,''
Phys. Lett. B \textbf{833} (2022), 137346
[arXiv:2112.12779 [hep-ph]].

\bibitem{Fang:2024hzy}
D.~L.~Fang, Y.~F.~Li, Y.~Y.~Zhang and J.~Y.~Zhu,
``Neutrinoless double beta decay in the minimal type-I seesaw model: mass-dependent nuclear matrix element, current limits and future sensitivities,''
JHEP \textbf{08} (2024), 217
[arXiv:2404.12316 [hep-ph]].

\bibitem{Xing:2009ce}
Z.~z.~Xing,
``Low-energy limits on heavy Majorana neutrino masses from the neutrinoless double-beta decay and non-unitary neutrino mixing,''
Phys. Lett. B \textbf{679} (2009), 255-259
[arXiv:0907.3014 [hep-ph]].

\bibitem{Doi:1985dx}
M.~Doi, T.~Kotani and E.~Takasugi,
``Double beta Decay and Majorana Neutrino,''
Prog. Theor. Phys. Suppl. \textbf{83} (1985), 1

\bibitem{Xing:2022hst}
Z.~z.~Xing,
``The translational \ensuremath{\mu}-\ensuremath{\tau} reflection symmetry of Majorana neutrinos,''
Int. J. Mod. Phys. A \textbf{38} (2023) no.01, 2250215
[arXiv:2207.08568 [hep-ph]].

\bibitem{Han:2021qum}
H.~c.~Han and Z.~z.~Xing,
``A full parametrization of the $9 \times 9$ active-sterile flavor mixing matrix in
the inverse or linear seesaw scenario of massive neutrinos,''
Nucl. Phys. B \textbf{973} (2021), 115609
[arXiv:2110.12705 [hep-ph]].

\bibitem{Wang:2021wdq}
Y.~Wang, B.~Yu and S.~Zhou,
``Flavor invariants and renormalization-group equations in the leptonic sector with massive Majorana neutrinos,''
JHEP \textbf{09} (2021), 053
[arXiv:2107.06274 [hep-ph]].

\bibitem{Yu:2021cco}
B.~Yu and S.~Zhou,
``Hilbert series for leptonic flavor invariants in the minimal seesaw model,''
JHEP \textbf{10} (2021), 017
[arXiv:2107.11928 [hep-ph]].

\bibitem{Yu:2022nxj}
B.~Yu and S.~Zhou,
``Spelling out leptonic CP violation in the language of invariant theory,''
Phys. Rev. D \textbf{106} (2022) no.5, L051701
[arXiv:2203.00574 [hep-ph]].

\bibitem{Yu:2022ttm}
B.~Yu and S.~Zhou,
``CP violation and flavor invariants in the seesaw effective field theory,''
JHEP \textbf{08} (2022), 017
[arXiv:2203.10121 [hep-ph]].

\end{thebibliography}
\end{document}